\documentstyle[preprint,prd,aps,eqsecnum]{revtex}

\begin{document}

\title{The Zel'dovich-type approximation for an inhomogeneous universe
in general relativity: second-order solutions}

\author{
Heinz Russ$^{1}$, Masaaki Morita$^{2}$,  Masumi Kasai$^{2,3}$
and Gerhard B\"orner$^{3}$}

\address{\small\it
${^1}$ Institut f\"{u}r Astronomie und Astrophysik,
Universit\"{a}t T\"{u}bingen, Auf der Morgenstelle 10,
72076 T\"{u}bingen, Germany
  \\
${^2}$ Department of Physics, Hirosaki University, 3 Bunkyo-cho,
Hirosaki, 036 Japan
  \\
${^3}$ Max-Planck Institut f\"ur Astrophysik, Karl-Schwarzschild
Str. 1, 85740 Garching, Germany
}

\date{December 11, 1995}

\maketitle


\begin{abstract}

The gravitational instability of inhomogeneities
in the expanding universe is studied
by the relativistic second-order approximation.
Using the tetrad formalism we consider irrotational dust universes
and get equations very similar to those given in the Lagrangian
perturbation theory in Newtonian cosmology.
Neglecting the cosmological constant and assuming a flat background model
we give the solutions of the nonlinear dynamics
of cosmological perturbations.
We present the complete second-order solutions,
which extend and improve earlier works.

\end{abstract}


\section{Introduction} \label{sec:intro}

Gravitational instability and structure formation in the universe
is an important topic of cosmological research.
By using N-Body codes it is possible
to follow the general nonlinear evolution
of initially small perturbations numerically,
but an understanding of what has happened between input and output
can often better be gained by analytical treatments.
Various analytical approaches have been compared
with the numerical results statistically \cite{bi:coles,bi:melott}
and it has turned out that the celebrated Zel'dovich approximation
\cite{bi:zeldo} gives the best fit to the numerical treatment.
Buchert \cite{bi:buch2} presented the Lagrangian perturbative
approximation to first-order based on Newtonian theory.
This work was extended to second-order \cite{bi:bueh} and
even to  third-order \cite{bi:buch4},
giving some new useful information about self-gravitating systems.
But these and most other analytical treatments are Newtonian approaches,
which are valid only for perturbations on scales much smaller than the
horizon size. On super-horizon scales instead one needs a relativisic
approach. The pioneer was Lifshitz \cite{bi:lifshitz46} with his
linearized theory on the basis of general relativity, which was
extended to the second-order by Tomita
\cite{bi:tomita67,bi:tomita71,bi:tomita72}.
A Lagrangian relativistic approximation to the second-order
based on fluid flow equations
was given by Matarrese, Pantano and Saez \cite{bi:mata2,bi:mata}.
Parry, Salopek and Stewart \cite{bi:pss} presented the nonlinear solution
of the Hamilton-Jacobi equation for general relativity,
using the spatial gradient expansion technique \cite{bi:ss}
and reproduced the Zel'dovich approximation.
The ``Higher-order Zel'dovich approximation'' is discussed in
Croudace et al. \cite{bi:cpss} and Salopek, Stewart and Croudace
\cite{bi:ssc}.

In this paper, we give an alternative approach, which extends
a tetrad-based Zel'dovich-type approximation
by Kasai \cite{bi:kasai95}. We derive the fully general relativistic
equations very similar to those given in the Newtonian case,
which are solved in a flat background model
without the cosmological constant by an iteration method.
The complete solutions are compared with previous work
and it is found that they include all these results.

This paper is organized as follows.
In Sec.~II we present the basic relativistic equations and
introduce the tetrad formalism.
In Sec.~III the perturbative approach is presented
and the solutions up to second-order are given.
In Sec.~IV we compare our results with previous works.
Sec.~V contains conclusions.
In the appendices we explain our gauge condition and present the
complete second-order solutions,
including the decaying and the coupling mode.
Units are chosen so that $c=1$.
Indices $\mu, \nu, \cdots$ and $a, b, \cdots$ run from $0$ to $3$
and indices $i, j, \cdots$ run from $1$ to $3$.

\section{Exposition of the Method}

In this section, we summarize a general relativistic treatment to
describe the non-linear evolution of an inhomogeneous universe
\cite{bi:kasai95,bi:kasai1,bi:kasai93}.  The models we consider contain
irrotational dust with density $\rho$ and four-velocity $u^{\mu}$ (and
possibly a cosmological constant $\Lambda$).
Neglecting the fluid pressure and the vorticity is a reasonable assumption
in a cosmological context.  In comoving synchronous coordinates, the
line element can be written in the form
\begin{equation}
ds^2 = -dt^2 + g_{ij} dx^i dx^j
\end{equation}
with $u^{\mu}=(1,0,0,0)$.  Then the Einstein equations read
\begin{eqnarray}
\label{eq:G00}
\frac{1}{2} \left[\;{^3\!R^i_{\ i}} + \left( K^i_{\ i} \right)^2
                          - K^i_{\ j} K^j_{\ i}\;\right]
  &=& 8\pi G\rho+\Lambda \;,\\
\label{eq:G0i}
K^i_{\ j \|i} - K^i_{\ i \|j}&=&0 \;,\\
\label{eq:Rij}
\dot{K}^i_{\ j} + K^{k}_{\ k} K^i_{\ j} + {^3\!R}^i_{\ j}
  &=& (4\pi G\rho+\Lambda)\, \delta^i_{\ j} \;,
\end{eqnarray}
where ${^3\!R}^i_{\ j}$ is the three-dimensional Ricci-tensor,
\begin{equation}\label{excurv}
K^i_{\ j} = \frac{1}{2} g^{ik}\dot g_{jk}
\end{equation}
is the extrinsic curvature, $\|$ denotes the covariant derivative with
respect to the three metric $g_{ij}$, and an overdot ($\dot{}$)
denotes $\partial/\partial t$.
The energy equation $u_{\mu}T^{\mu\nu}_{\ \ ;\nu}=0$ gives
\begin{equation}\label{ka3}
\dot\rho+\rho K^i_{\ i} = 0 \;,
\end{equation}
with the solution
\begin{equation}
\label{eq:rho}
\rho = \rho(t_{in},\mbox{\bf x})
  \frac{\sqrt{\mbox{\rm det} \left[ g_{ij}(t_{in},\mbox{\bf x}) \right] }}
       {\sqrt{\mbox{\rm det} \left[ g_{ij}(t,     \mbox{\bf x}) \right] }}\; .
\end{equation}
The evolution equation for the Ricci curvature is obtained in the
form\cite{bi:kasai95}
\begin{equation}\label{bianchi}
{^3\!\dot R}^i_{\ j} + 2 K^i_{\ k}\,{^3\!R}^{k}_{\ j} =
  K^{i\ \ \ \|k}_{\ k \| j}
+ K^{k\ \|i}_{\ j \ \ \| k}
- K^{i\ \|k}_{\ j \ \ \| k}
- K^{k\ \|i}_{\ k \ \ \| j} \;.
\end{equation}

Let us introduce the scale factor function $a(t)$ which satisfies the
following equation
\begin{equation}\label{eq:ddota}
2 a \ddot{a} + \dot{a}^2 + k - \Lambda a^2 = 0 \;,
\end{equation}
where the curvature constant $k$ takes the value of $+1, 0, -1$ for
closed, flat, and open spaces, respectively.
(Eq.~(\ref{eq:ddota}) is obtained from the Friedmann equation
\begin{equation}
\left(\frac{\dot a}{a}\right)^2 + \frac{k}{a^2} =
\frac{8\pi G}{3} \rho_b + \frac{\Lambda}{3}
\end{equation}
and its derivative with respect to time.)
If the spacetime is exactly Friedmann-Lema\^{\i}tre-Robertson-Walker
(FLRW), then we have
\begin{equation}
K^i_{\ j} = \frac{\dot a}{a} \delta^i_{\ j}, \qquad
{^3\!R}^i_{\ j} = 2 \frac{k}{a^2}\delta^i_{\ j}
\qquad \mbox{for FLRW}\;.
\end{equation}
Therefore, the deviations from the FLRW models due to inhomogeneity
are expressed by the peculiar part of the extrinsic curvature
\begin{equation}
V^i_{\ j} \equiv K^i_{\ j} - \frac{\dot a}{a}\delta^i_{\ j} \;,
\end{equation}
which represents the deviation from the uniform Hubble expansion, and
the deviation of the curvature tensor
\begin{equation}
{\cal R}^i_{\ j} \equiv a^2 \; {^3\!R}^i_{\ j} - 2 k \delta^i_{\ j}
 \;.
\end{equation}
Using these quantities, Eqs.~(\ref{eq:G0i}), (\ref{eq:Rij}), and
(\ref{bianchi}) are rewritten as
\begin{equation}
\label{eq:Vij}
V^i_{\ j | i} - V^i_{\ i | j} = 0 \;,
\end{equation}
\begin{equation}
\label{eq:dotVij}
\dot{V}^i_{\ j} +
  \left( 3 {\dot a \over a} + V^k_{\ k} \right) V^i_{\ j}
  + {1 \over {a^2}}
    \left( {\cal R}^i_{\ j} -
           {1 \over 4} {\cal R}^k_{\ k} \,\delta^i_{\ j} \right) =
  {1 \over 4} \left\{ (V^k_{\ k})^2 - V^k_{\ \ell} V^{\ell}_{\ k} \right\}
  \delta^i_{\ j} \;,
\end{equation}
\begin{equation}
\label{eq:dotRR}
\dot{{\cal R}}^i_{\ j} + 2 V^i_{\ \ell} \,{\cal R}^{\ell}_{\ j}
 + 4 k\, V^i_{\ j}
  = V^{i\ \ |\ell}_{\ \ell | j}
  + V^{\ell \ | i}_{\ j \ \ | \ell}
  - V^{i \ | \ell}_{\ j \ \ | \ell}
  - V^{\ell \ | i}_{\ \ell \ \ | j} \; ,
\end{equation}
where \(|\) denotes the covariant derivative with respect to the
conformally transformed three metric \(\gamma_{ij} \equiv a^{-2}
g_{ij}\).

The procedure essential to develop the relativistic Zel'dovich-type
approximation\cite{bi:kasai95} is to introduce the following orthonormal
tetrad
\begin{equation}
g_{\mu\nu}=\eta_{(a)(b)}\bar e^{(a)}_{\ \mu}\bar e^{(b)}_{\ \nu}
\end{equation}
with
\begin{equation}
  \bar{e}^{(0)}_{\ \mu} = u_{\mu} = (-1, 0, 0, 0) \;, \qquad
  \bar{e}^{(\ell)}_{\ \mu}
  =      \left( 0, \bar{e}^{(\ell)}_{\ i} \right)
  \equiv \left( 0, a(t)\,  e^{(\ell)}_{\ i} \right) \quad
\mbox{for $\ell = 1, 2, 3$.}
\end{equation}
The spatial basis vectors are parallelly transported along each fluid
line, i.e.,
\begin{equation}
\bar{e}^{(\ell)}_{\ \mu; \nu}\, u^{\nu} = 0 \;.
\end{equation}
In our choice of the tetrad components, it reads
\begin{equation}
\label{eq:dote}
\dot{e}^{(\ell)}_{\ i} = V^j_{\ i}\, e^{(\ell)}_{\ j}
\qquad  \mbox{or} \qquad
V^i_{\ j} = e^i_{(\ell)}\,\dot{e}^{(\ell)}_{\ j} \;.
\end{equation}
Using Eqs.~(\ref{eq:dotVij}), (\ref{eq:dotRR}) and (\ref{eq:dote}), we
obtain the following key equation
\begin{equation}\label{eq:key}
\frac{\partial}{\partial t}
\left[a^3
  \left(\ddot e^{(\ell)}_{\ i} + 2\frac{\dot a}{a} \dot e^{(\ell)}_{\ i}
        -4\pi G\rho_b e^{(\ell)}_{\ i}
  \right)
\right]
= a \left(P^{(\ell)}_{\ i} + Q^{(\ell)}_{\ i} + S^{(\ell)}_{\ i}\right) \;,
\end{equation}
where
\begin{equation}
P^{(\ell)}_{\ i} = \frac{\partial}{\partial t}
\left\{
    \frac{a^2}{4} \left[
       \left(V^k_{\ k}\right)^2 - V^k_{\ n} V^n_{\ k}
                   \right] e^{(\ell)}_{\ i}
  - a^2 \left(V^k_{\ k} V^j_{\ i} -  V^j_{\ k} V^k_{\ i} \right)
  e^{(\ell)}_{\ j}
\right\} \;,
\end{equation}
\begin{equation}
Q^{(\ell)}_{\ i}  =
 \left( V^j_{\ k} {\cal R}^k_{\ i}
       +\frac{1}{4} V^j_{\ i} {\cal R}^k_{\ k}
       -\frac{1}{2} \delta^j_{\ i} \, V^k_{\ n}{\cal R}^n_{\ k}
   \right) e^{(\ell)}_{\ j}
\;,
\end{equation}
and
\begin{equation}
S^{(\ell)}_{\ i} =
  \left[\, V^{j \ | k}_{\ i \ \ | k}
          +V^{k \ | j}_{\ k \ \ | i}
          -V^{j \ \ | k}_{\ k | i  }
          -V^{k \ | j}_{\ i \ \ | k}
          +k \left( 3V^j_{\ i} - V^k_{\ k} \delta^j_{\ i} \right)
  \,\right] e^{(\ell)}_{\ j} \;.
\end{equation}
Note that the left-hand-side of Eq.~(\ref{eq:key}) is already
linearized with respect to \( e^{(\ell)}_{\ i} \), and all terms on the
right-hand-side, except \(S^{(\ell)}_{\ i}\), are manifestly nonlinear
quantities.
It has, therefore, a form suitable for solving it perturbatively by iteration.
It should also be stressed that we have not used any approximation
methods in deriving Eq.~(\ref{eq:key}). Our treatment here is fully
nonlinear and exact.

\section{Perturbative approach}

In this section, we solve perturbatively the key equation
Eq.~(\ref{eq:key}) by an iteration method.

\subsection{The background}

The background ( \( V^i_{\ j} = 0, {\cal R}^i_{\ j} = 0\) )
solution is characterized by
\begin{equation}
\dot{e}^{(\ell)}_{\ i} = 0 \;, \quad \mbox{i.e.,} \quad
 e^{(\ell)}_{\ i} = e^{(\ell)}_{\ i}(\mbox{\bf x}) \;.
\end{equation}
Furthermore, the metric
\( \gamma_{ij} = \delta_{(k)(\ell)} e^{(k)}_{\ i}
                                       e^{(\ell)}_{\ j} \)
is that of a constant curvature space
with curvature constant $k$.
In the case of a flat background, we can write
\begin{equation}
e^{(\ell)}_{\ i} = \delta^{(\ell)}_{\ i} \quad \mbox{for}\ \ k = 0 \;.
\end{equation}
Hereafter, we restrict our consideration to the Einstein-de Sitter
background, \(k = 0, \Lambda = 0\).

\subsection{The first-order solutions: scalar modes}

Linear perturbations are classified into scalar, vector, and
tensor modes. In the first-order level, they do not couple with each
other, and can be discussed separately. Let us first consider the
scalar perturbations.
The general form for the linearly perturbed triad in this case is
\begin{equation}
e^{(\ell)}_{\ i} = \delta^{(\ell)}_{\ i} + E^{(\ell)}_{\ i} =
  \delta^{(\ell)}_{\ i} + \delta^{(\ell)}_{\ j} \left(
                       A\, \delta^j_{\ i} + B^{,j}_{\ ,i} \right) \;.
\end{equation}

Let us write the first-order quantities with subscript $(1)$.
Then the perturbed extrinsic curvature is
\begin{equation}
V^i_{(1)\,j} = \dot{A}\,\delta^{i}_{\ j} +
               \dot{B}^{,i}_{\ ,j} \;.
\end{equation}
 From the constraint equation (\ref{eq:Vij}), which reads
\begin{equation}\label{eq:Vij1st}
V^i_{(1)\; j,i} - V^i_{(1)\; i,j} = 0
\end{equation}
in the first-order, we obtain \( \dot{A}_{,i} = 0 \).
However, the part \( \dot{A}(t)\,
\delta^i_{\ j} \) in the extrinsic curvature simply represents the
uniform and isotropic Hubble expansion.  Therefore, by a suitable
re-definition of the background, we can set
\begin{equation}
\dot{A} = 0 \;, \quad \mbox{i.e.,} \quad A = A(\mbox{\bf x}) \;.
\end{equation}

As was noted previously, it is apparent that the source terms $
P^{(\ell)}_{\ i} $ and $Q^{(\ell)}_{\ i} $ are second-order quantities
(and higher).
Using \(V^i_{(1)\, j} = \dot{B}^{,i}_{\ ,j}\), we also find that \(
S^{(\ell)}_{\ i} \) vanishes in linear order:
\begin{equation}
S^{(\ell)}_{(1)\, i} = \delta^{(\ell)}_{\ j} \left(
         V^{j \ \ ,k}_{(1)\, i \ ,k}
        +V^{k \ \ ,j}_{(1)\, k \ ,i}
        -V^{j \ \ \ \, ,k}_{(1)\, k ,i}
        -V^{k \ \ ,j}_{(1)\, i \ ,k} \right) = 0 \;.
\end{equation}
Therefore, to first-order, the right-hand-side of the key equation
(\ref{eq:key}) vanishes and it can be integrated to give
\begin{equation}\label{eq:ddotE}
a^3\left( \ddot E^{(\ell)}_{\ i}
         + 2\frac{\dot a}{a} \dot E^{(\ell)}_{\ i}
         - 4\pi G\rho_b E^{(\ell)}_{\ i}
   \right) = C^{(\ell)}_{\ i}(\mbox{\bf x}) \;.
\end{equation}
By choosing 
$C^{(\ell)}_{\ i}(\mbox{\bf x}) =-4\pi G\rho_b a^3 \,\delta^{(\ell)}_{\ j}
  \left(A(\mbox{\bf x}) \,\delta^j_{\ i} + C^{,j}_{\ ,i}(\mbox{\bf x})
\right)$,
Eq.~(\ref{eq:ddotE}) is re-written as
\begin{equation}
\frac{\partial^2}{\partial t^2}
  \left(B^{,i}_{\ ,j} - C^{,i}_{\ ,j}\right)
+ 2 \frac{\dot a}{a} \frac{\partial}{\partial t}
  \left(B^{,i}_{\ ,j} - C^{,i}_{\ ,j}\right)
- 4 \pi G \rho_b\, \left(B^{,i}_{\ ,j} - C^{,i}_{\ ,j}\right) = 0 \;.
\end{equation}
Note that now it has the same form as the equation which governs the density
contrast $\delta$ in conventional linear perturbation theory
\cite{bi:peebles}.
Using the growing mode
\(D^+(t) = a(t) = t^{2/3}\) and the decaying mode solutions
 \(D^-(t) = t^{-1}\) respectively,
we obtain the solutions in the form
\begin{equation}
B^{,i}_{\ ,j} = C^{,i}_{\ ,j}(\mbox{\bf x})
               + t^{\frac{2}{3}}\, \Psi^{,i}_{\ ,j}(\mbox{\bf x})
               + t^{-1 }\, \Phi^{,i}_{\ ,j}(\mbox{\bf x}) \;.
\end{equation}
For the metric, we have the following first-order expression:
\begin{equation}\label{eq:gij1st}
g_{ij} = a^2(t) \left[
         \left(1 + 2 A(\mbox{\bf x})\right)\delta_{ij}
        + 2 C_{,ij}(\mbox{\bf x})
        + 2 t^{\frac{2}{3}}\, \Psi_{,ij}(\mbox{\bf x})
        + 2 t^{-1 }\, \Phi_{,ij}(\mbox{\bf x}) \right] \;.
\end{equation}

The relation between $A(\mbox{\bf x})$ and $\Psi(\mbox{\bf x})$
is given by Eq.~(\ref{eq:dotVij}).
To first-order, it reads
\begin{equation}\label{eq:dotVij1st}
\dot V^i_{(1)\,j} + 3\frac{\dot a}{a}V^i_{(1)\,j}
+ \frac{1}{a^2} \left(
    {\cal R}^i_{(1)\,j}-\frac{1}{4}{\cal R}^k_{(1)\,k}\delta^i_{\ j}
                \right) = 0\;,
\end{equation}
where
\begin{equation}
V^i_{(1)\,j} = \frac{2}{3} t^{-\frac{1}{3}} \Psi^{,i}_{\ ,j}
              - t^{-2} \Phi^{,i}_{\ ,j}
\end{equation}
and
\begin{equation}
{\cal R}^i_{(1)\,j} = - A^{,i}_{\ ,j} - A^{,k}_{\ ,k}\,\delta^i_{\ j}
\;.
\end{equation}
Hence we have
\begin{equation}
A(\mbox{\bf x}) = \frac{10}{9} \Psi(\mbox{\bf x}) \;.
\end{equation}
The function \(C(\mbox{\bf x})\) is not determined by the Einstein
equations within our approximation. As shown in APPENDIX A, however, we
can set $C(\mbox{\bf x}) = 0$ using a residual gauge freedom.
The final form of the first-order solutions is, therefore,
\begin{equation}
e^{(\ell)}_{\ i} =
  \left( 1 + \frac{10}{9} \Psi(\mbox{\bf x}) \right)
    \delta^{(\ell)}_{\ i} +
  \delta^{(\ell)}_{\ j} \left(
                 t^{\frac{2}{3}}\, \Psi^{,j}_{\ ,i}(\mbox{\bf x})
               + t^{-1 }\, \Phi^{,j}_{\ ,i}(\mbox{\bf x})
                        \right) \;,
\end{equation}
or in the form of the metric
\begin{equation}
g_{ij} = a^2(t) \left[
         \left(1 + \frac{20}{9}\Psi(\mbox{\bf x})\right)\delta_{ij}
        + 2 t^{\frac{2}{3}}\, \Psi_{,ij}(\mbox{\bf x})
        + 2 t^{-1 }\, \Phi_{,ij}(\mbox{\bf x}) \right] \;.
\end{equation}
Note that we have not assumed that the density contrast
is small, in order to derive the solutions.
The density is given by Eq.~(\ref{eq:rho}), which in this case reads
\begin{equation}
\rho = \rho(t_{in},\mbox{\bf x})
       \left(\frac{a(t_{in})}{a(t)}\right)^3
       \frac{\mbox{\rm det}[e^{(\ell)}_{\ i}(t_{in},\mbox{\bf x})]}
            {\mbox{\rm det}[e^{(\ell)}_{\ i}(t,     \mbox{\bf x})]} \;.
\end{equation}

\subsection{The first-order solutions: tensor modes}

Under the assumption of vanishing vorticity, the remaining is the
tensor mode perturbations. In this case, we can write the triad in the
form
\begin{equation}\label{eq:1tensor}
e^{(\ell)}_{\ i} = \delta^{(\ell)}_{\ i} +
                   \delta^{(\ell)}_{\ j} H^j_{\ i}
\end{equation}
with \(\  H^i_{\ j,i} = 0 \ \) and \(\  H^i_{\ i} = 0\).

The perturbed extrinsic curvature is
\begin{equation}
V^i_{(1)j} = \dot{H}^i_{\ j} \; .
\end{equation}
Then, the constraint equation (\ref{eq:Vij}) to first-order,
i.e., Eq.~(\ref{eq:Vij1st}), is trivially satisfied.

To obtain the equation for $H^i_{\ j}$, we can use the key equation
(\ref{eq:key}).
On the right-hand-side of Eq.~(\ref{eq:key}),
$S^{(\ell)}_{\ i}$ is the only quantity to be calculated
since $P^{(\ell)}_{\ i}$ and $Q^{(\ell)}_{\ i}$ are of higher order:
\begin{equation}
S^{(\ell)}_{(1)\, i}
        = \delta^{(\ell)}_{\ j} \dot{H}^{j\ ,k}_{\ i\ ,k} \; .
\end{equation}
Eq.~(\ref{eq:key}) reads
\begin{equation}\label{eq:key1stTT}
\frac{\partial}{\partial t}
\left[\, a^3
  \left(\ddot H^i_{\ j} + 2\frac{\dot a}{a} \dot H^i_{\ j}
        -4\pi G\rho_b H^i_{\ j}
  \right)
\, \right] = a\,\dot{H}^{i\ ,k}_{\ j\ ,k} \; .
\end{equation}
Integrating Eq.~(\ref{eq:key1stTT}), we obtain
\begin{equation}\label{eq:ddotH}
\ddot H^i_{\ j} + 3\frac{\dot a}{a} \dot H^i_{\ j}
                - \frac{1}{a^2}\, \nabla^2 H^i_{\ j} = 0 \; ,
\end{equation}
where $\nabla^2$ is the Laplacian of flat 3-spaces.
In fact, the same equation for $H^i_{\ j}$ can be also obtained
directly from Eq.~(\ref{eq:dotVij1st}).
The solution of Eq.~(\ref{eq:ddotH}) is given as
\begin{equation}
  H^i_{\ j} = \int d^3 \mbox{\bf q} \ \; t^{-\frac12}
              J_{\pm\frac32}(3|\mbox{\bf q}|t^{\frac13}) \ \,
              h^i_{\ j} \exp(i \mbox{\bf q} \cdot \mbox{\bf x}) \; ,
\end{equation}
where $J_{\pm\frac32}$ is the Bessel function of order $\pm 3/2$
and $h^i_{\ j}$ is a constant tensor with
$ h^i_{\ i} = 0 $ and $ q_i h^i_{\ j} = 0 $.
(See, e.g., Ref. \cite{bi:weinberg} for detail.)

\subsection{The second-order solutions}

In order to avoid notational complexity, in this subsection we only
deal with growing mode terms. The complete solutions of the decaying
and coupling terms can be found in APPENDIX B.
Moreover, we omit the first-order tensor mode.
(It is not our aim to consider the nonlinear effect
which comes from this mode.
With respect to this problem, see Ref. \cite{bi:tomita71,bi:tomita72}.)
Thus we begin with the following form
\begin{equation}\label{eq:e2nd}
e^{(\ell)}_{\ i} = \left( 1 + \frac{10}{9} \Psi(\mbox{\bf x}) \right)
    \delta^{(\ell)}_{\ i} +
   t^{\frac{2}{3}}\,  \delta^{(\ell)}_{\ j}  \Psi^{,j}_{\ ,i}(\mbox{\bf x})
 + \varepsilon^{(\ell)}_{\ i} \;.
\end{equation}
The second-order quantity $\varepsilon^{(\ell)}_{\ i}$ is decomposed into
a transverse-traceless part and a remaining longitudinal
part
\begin{equation}
\varepsilon^{(\ell)}_{\ i} = \delta^{(\ell)}_{\ j} \left(
     \beta^j_{\ i} + \chi^j_{\ i}                 \right) \;,
\end{equation}
where $\ \chi^i_{\ j,i} = 0, \ \chi^i_{\ i}= 0$.

The peculiar deformation tensor to second-order is immediately
found to give
\begin{equation}\label{eq:pecdef2}
V^i_{(2)\, j} = \dot\beta^i_{\ j} + \dot\chi^i_{\ j}
           -\frac{20}{27} t^{-\frac{1}{3}} \Psi\Psi^{,i}_{\ ,j}
           -\frac{2}{3}t^{\frac{1}{3}}\Psi^{,i}_{\ ,k}\Psi^{,k}_{\ ,j}\;.
\end{equation}
Quantities with subscript $(2)$ represent the second-order term in the
expansion.
 From the constraint equation (\ref{eq:Vij}), we now obtain
\begin{equation}\label{eq:dotbetaij}
\dot\beta^i_{\ j,i} - \dot\beta^i_{\ i,j} +
\frac{20}{27} t^{-\frac{1}{3}} \left( \Psi^{,k} \Psi_{,k} \right)_{,j}
 = 0 \;.
\end{equation}

Let us turn our attention to the key equation (\ref{eq:key}).
To second-order it reads
\begin{equation}\label{eq:key2nd}
\ddot\varepsilon^{(\ell)}_{\ i}
  + 2 \frac{\dot a}{a} \dot\varepsilon^{(\ell)}_{\ i}
  - 4 \pi G \rho_b \varepsilon^{(\ell)}_{\ i}
  =
    \frac{1}{a^3} c^{(\ell)}_{\ i}(\mbox{\bf x})
  + \frac{1}{a^3} \int^t a\,\left( P^{(\ell)}_{(2)\, i}
                                  +Q^{(\ell)}_{(2)\, i}
                                  +S^{(\ell)}_{(2)\, i}\right)dt \;,
\end{equation}
where $c^{(\ell)}_{\ i}(\mbox{\bf x})$ is a second-order integration
``constant''.
It is apparent that the source terms $P^{(\ell)}_{(2)\, i}$ and
$Q^{(\ell)}_{(2)\, i}$ are quadratic with respect to the first-order
quantities, hence contain neither $\beta^i_{\ j}$ nor $\chi^i_{\ j}$.
Furthermore, from Eq.~(\ref{eq:dotbetaij}), we find that the
longitudinal part of $S^{(\ell)}_{(2)\, i}$ does not contain
$\beta^i_{\ j}$.
Actually, if we take the divergence of Eq.~(\ref{eq:key2nd}),
we obtain
\begin{equation}
  \ddot\beta^i_{\ j,i}
    + 2 \frac{\dot a}{a} \dot\beta^i_{\ j,i}
    - 4 \pi G \rho_b \beta^i_{\ j,i}
   =
      \frac{1}{a^3} c^i_{\ j,i}(\mbox{\bf x})
    - \frac13 t^{-\frac23} \left(
       (\Psi^{,k}_{\ ,k})^2 - \Psi^{,k}_{\ ,\ell}\Psi^{,\ell}_{\ ,k}
                          \right)_{,j}.
\end{equation}
Therefore, solutions for $\beta^i_{\ j}$ can be written as a
linear combination of the homogeneous solution and  the inhomogeneous
solution in the presence of the given source terms:
\begin{equation}
\beta^i_{\ j} = \alpha(\mbox{\bf x})\,\delta^i_{\ j}
  + t^{\frac23} \, \psi^i_{\ j} (\mbox{\bf x})
  + t^{\frac43} \, \varphi^i_{\ j} (\mbox{\bf x}) \;,
\end{equation}
where we have used a convenient choice of the integration
``constant'', $c^{(\ell)}_{\ i}(\mbox{\bf x}) =
 - 4 \pi G \rho_b a^3 \alpha(\mbox{\bf x})\,\delta^i_{\ j}$.

Once we obtain the temporal dependency of the solutions, their spatial
dependency, i.e., $\psi^i_{\ j} (\mbox{\bf x})$ and $\varphi^i_{\ j}
(\mbox{\bf x})$ are determined by Eq.~(\ref{eq:dotVij}).
To second-order
\begin{eqnarray}
\psi^i_{\ j} &=& \frac{5}{9} \Psi^{,k} \Psi_{,k} \delta^i_{\ j}
               - \frac{10}{9} \left(\Psi^2\right)\!{}^{,i}_{\ ,j}
               + \frac{9}{10} \alpha^{,i}_{\ ,j} \;, \\
\varphi^i_{\ j} &=& \frac{3}{7} \left( \mu^k_{\ k} \,\delta^i_{\ j}
                   -4 \mu^i_{\ j} \right) \;,
\end{eqnarray}
where
\begin{equation}
\mu^i_{\ j} \equiv \frac{1}{2} \left(
  \Psi^{,k}_{\ ,k} \Psi^{,i}_{\ ,j} -
  \Psi^{,i}_{\ ,k} \Psi^{,k}_{\ ,j}\right) \;.
\end{equation}
(The tensor $\mu^i_{\ j}$ has an interesting property:
the trace $\mu^i_{\ i}$ gives the second scalar invariant
\footnote{
Three scalar invariants of a three-dimensional tensor $A^i_{\ j}$ are
defined by
$I(A) \equiv A^i_{\ i}$,
$I\!I(A) \equiv 1/2 [(A^i_{\ i})^2 - A^i_{\ j} A^j_{\ i}]$,
$I\!I\!I(A) \equiv \mbox{det}(A^i_{\ j})$.
They satisfy the relation
$\mbox{\rm det}(\delta^i_{\ j} + A^i_{\ j}) = 1 + I(A) +
I\!I(A) + I\!I\!I(A)$. See, e.g., Ref.~\cite{bi:buch4} and
references therein.} of the tensor $\Psi^{,i}_{\ ,j}$. )

The equation for $\chi^i_{\ j}$ can also be obtained from
Eq.~(\ref{eq:key2nd}), but it is more convenient to use
Eq.~(\ref{eq:dotVij}) instead.  To second-order, it gives for the
transverse-traceless part
\begin{equation}\label{eq:ddotchi}
 \ddot\chi^{i}_{\ j} + 3\frac{\dot a}{a}\dot\chi^{i}_{\ j}
    - \frac{1}{a^2}\nabla^2\chi^{i}_{\ j}
  = {\cal S}^{i}_{\ j} \;,
\end{equation}
where
\begin{equation}
{\cal S}^{i}_{\ j} =
   \frac{3}{7} \mu^{k\ ,i}_{\ k\ ,j}
 + \frac{3}{7} \left(\mu^{k}_{\ k}\, \delta^i_{\ j} - 4 \mu^i_{\ j}
              \right)\!{}^{,\ell}_{\ ,\ell}
\end{equation}
is a transverse and traceless tensor:
${\cal S}^i_{\ i} = 0, \ \ {\cal S}^i_{\ j,i} = 0$.
This shows that gravitational waves are induced even if there are
initially scalar perturbations only.
The solution of Eq.~(\ref{eq:ddotchi}) was given by Tomita
\cite{bi:tomita67} in the following way.
Introducing the conformal time variable $\eta$,
which is related to $t$ by $dt = a d\eta$,
Eq.~(\ref{eq:ddotchi}) is rewritten as
\begin{equation}\label{ddetachi}
 \frac{\partial^2}{\partial\eta^2} \chi^i_{\ j}
    + \frac{4}{\eta} \frac{\partial}{d\eta} \chi^i_{\ j}
    - \nabla^2 \chi^i_{\ j}
  = \frac{1}{81} \eta^4 {\cal S}^{i}_{\ j} \; .
\end{equation}
Eq.~(\ref{ddetachi}) can be solved
using the retarded Green function as
\begin{equation}\label{solchi}
 \chi^i_{\ j}(\mbox{\bf x},\eta)
   = \frac{1}{81}
        \int D_{\mbox{ret}}(\mbox{\bf x},\eta \, ; \mbox{\bf x'},\eta')
         \  a^4(\eta') \  {\cal S}^i_{\ j}(\mbox{\bf x'})
                \ d\eta' \, d^3\mbox{\bf x'} \; ,
\end{equation}
where
\begin{equation}\label{retard}
  D_{\mbox{ret}}(\mbox{\bf x},\eta \, ; \mbox{\bf x'},\eta')
     = \frac{1}{4\pi(\eta\eta')^3}\left(1+\epsilon(\eta-\eta')\right)
        \left(
        \eta \eta' \delta \left(\tau^2-r^2\right)
      + \frac{1}{2} \theta \left(\tau^2-r^2\right)
        \right)
\end{equation}
with $r\equiv|\mbox{\bf x}-\mbox{\bf x'}|$ and $\tau \equiv \eta - \eta'$.
Substituting Eq.~(\ref{retard}) into Eq.~(\ref{solchi}),
the solution reads
\begin{equation}\label{chisolution}
 \chi^i_{\ j}(\mbox{\bf x},\eta)
   = \frac{1}{1944\pi\eta^3}
       \int_0^{\eta - \eta_{in}}  dr'\ r' \left(
            (6\eta + r')(\eta - r')^6 - r'\eta_{in}^{\ 6}
                                              \right)
        \int d\Omega'\ {\cal S}^i_{\ j}(\mbox{\bf x} + \mbox{\bf x'})\; ,
\end{equation}
where $r'\equiv |\mbox{\bf x'}|$.

Finally, we obtain the metric tensor up to second-order
\begin{eqnarray}
\gamma_{ij} = & &  a^{-2}\,g_{ij} =
  \left(1 + \frac{20}{9}\Psi + \frac{100}{81}\Psi^2 + 2 \alpha \right)
    \delta_{ij} \nonumber\\
 && + \,a(t)\left[ \left( 2\Psi - \frac{20}{9}\Psi^2 + \frac{9}{5}\alpha
                 \right)\!{}_{,ij}
                +\frac{20}{9}\Psi\Psi_{,ij}
                +\frac{10}{9}\Psi^{,k}\Psi_{,k}\,\delta_{ij}
          \right] \\
 && + \,a^2(t)\left[ \frac{19}{7} \Psi^{,k}_{\ ,i}\Psi_{,kj}
                  -\frac{12}{7} \Psi^{,k}_{\ ,k}\Psi_{,ij}
                  +\frac{3}{7}\left( \left(\Psi^{,k}_{\ ,k}\right)^2
                    - \Psi^{,k}_{\ ,\ell}\Psi^{,\ell}_{\ ,k} \right)
                   \delta_{ij}
            \right] + 2 \chi_{ij}  \;. \nonumber
\end{eqnarray}
We still have freedom in choosing $\alpha(\mbox{\bf x})$,
which corresponds to the second-order term of the initial amplitude of the
gravitational potential fluctuations.
It can be absorbed into the first-order perturbations by a suitable
re-definition of $\Psi(\mbox{\bf x})$.
For example, choosing $\alpha = -\frac{50}{81} \,\Psi^2$ gives
\begin{eqnarray}\label{eq:gammaij2nd2}
\gamma_{ij} = && \left(1 + \frac{20}{9}\Psi \right) \delta_{ij}
                  + 2 a(t) \Psi_{,ij} \nonumber\\
 && + \,\frac{10}{9} a(t)
          \left( -6 \Psi_{,i}\Psi_{,j} - 4\Psi\Psi_{,ij}
                + \Psi^{,k}\Psi_{,k}\,\delta_{ij}
          \right) \\
 && + \,\frac{1}{7} a^2(t)\left[ 19 \Psi^{,k}_{\ ,i}\Psi_{,kj}
                  -12 \Psi^{,k}_{\ ,k}\Psi_{,ij}
                  +3\left( \left(\Psi^{,k}_{\ ,k}\right)^2
                    - \Psi^{,k}_{\ ,\ell}\Psi^{,\ell}_{\ ,k} \right)
                   \delta_{ij}
            \right] + 2 \chi_{ij}  \;. \nonumber
\end{eqnarray}
At the initial time ($t \rightarrow 0$) only first-order
metric perturbations exist.

\section{Comparison with previous works}

In this section, we compare our result (Eq.~(\ref{eq:gammaij2nd2}))
with previous work. Quantities, which refer to these papers,
will be indicated by a hat.

\subsection{Tomita's second-order theory}

Tomita \cite{bi:tomita67} extended Lifshitz's linearized theory
\cite{bi:lifshitz46} up to the second-order calculation
on the basis of general relativity. Setting
$\hat F = \frac{20}{9} \Psi$ for the growing mode and
$\hat F = 54\Phi$ for the decaying mode
(see Eq.~(4.1) in Ref. \cite{bi:tomita67})
his result is fully coincident with ours, except for one point:
he did not consider the terms due to the coupling between the growing and
decaying modes, which are included in our complete solutions.
(See APPENDIX B.)

\subsection{The fluid flow approach}

Matarrese et al. \cite{bi:mata} also carried out second-order calculations
based on the fluid flow approach. Their result
(Eq.~(49) in Ref. \cite{bi:mata}) is partly consistent
with ours, since they neglect several terms in the computed metric.
In spite of the fact that they obtain the initial condition from the
gauge-invariant linear theory, they neglect the first-order constant mode,
$\frac{20}{9} \,\Psi \delta_{ij}$ in our notation
in Eq.~(\ref{eq:gammaij2nd2}) in the subsequent calculations.
Also missing is the second-order homogeneous solution,
which is proportional to $t^{2/3}$.

The comparison of the second-order transverse-traceless parts
has to be taken with some caution.
Eq.~(B19) in Ref. \cite{bi:mata}, which has to be solved,
can be derived from our Eq.~(\ref{eq:ddotchi}).
In the short-wavelength limit inside the horizon
$(\eta \gg r')$ in our approach we get
$\nabla^2 \chi^i_{\ j} = -t^{\frac{4}{3}}{\cal S}^i_{\ j}$,
which can be identified with Eq.~(65) in Ref. \cite{bi:mata}.
In the long-wavelength limit
outside the horizon they obained a result, which can be neglected
cause of the appearance of spatial derivatives,
whereas in our exact result
there exists no solution for the wavelength larger than the
horizon size.

\subsection{The gradient expansion technique}

Parry et al. \cite{bi:pss} derived a nonlinear solution for $g_{ij}$,
based on the gradient expansion method.
(See also Ref. \cite{bi:cpss,bi:ssc}.)
Their ``fifth-order'' result is the
\begin{eqnarray}
\label{eq:pss-5th}
  \hat \gamma_{ij} = && t^{\frac43} \hat k_{ij} + \frac{9}{20} t^2
               \left( \hat R \hat k_{ij} - 4\hat R_{ij} \right)
       + \frac{81}{350} t^{\frac83}
           \biggm[ \left(\;
         \frac{89}{32}\hat R^2 + \frac58 \hat R^{;k}_{\ ;k}
         - 4\hat R^{k\ell}\hat R_{k\ell}
                   \right) \hat k_{ij}    \nonumber\\
    && \hspace{3cm}
       - \, 10 \hat R \hat R_{ij} + \frac58 \hat R_{;ij}
         + 17 \hat R^{\ n}_i \hat R_{jn}
         - \frac52 \hat R^{\ \ \ ;k}_{ij;k}
           \biggm] \; ,
\end{eqnarray}
where $\hat k_{ij} = \hat k_{ij}(\mbox{\bf x})$ is the ``seed'' metric,
$\hat R_{ij}$ and $\hat R$ are the 3-dimensional Ricci tensor
and Ricci scalar, respectively, of the 3-metric $\hat k_{ij}$,
and a semicolon $(;)$ denotes the covariant derivative with respect to
$\hat k_{ij}$.
To compare our solution Eq.~(\ref{eq:gammaij2nd2}) with
their Eq.~(\ref{eq:pss-5th}), we set
\begin{equation}
  \hat k_{ij}(\mbox{\bf x})
          = \left( 1 + \frac{20}{9} \Psi(\mbox{\bf x})
            \right) \delta_{ij} \; .
\end{equation}
Then the Ricci tensor up to second-order is
\begin{eqnarray}
  \hat R_{ij} &=& -\frac{10}{9} \left(
                         \Psi_{,ij} + \Psi^{,k}_{\ ,k} \,\delta_{ij}
                                     \right)  \nonumber \\
              && + \,\frac{100}{27} \Psi_{,i} \Psi_{,j}
                       + \frac{200}{81} \Psi \Psi_{,ij}
                       + \left(
                           \frac{100}{81} \Psi^{,k} \Psi_{,k}
                         + \frac{200}{81} \Psi \Psi^{,k}_{\ ,k}
                         \right) \delta_{ij} \; .
\end{eqnarray}
If we substitute this expression into Eq.~(\ref{eq:pss-5th})
and calculate up to the second-order,
we obtain
\begin{eqnarray}
 \gamma_{ij} \equiv t^{-\frac43} \hat \gamma_{ij}
        = && \left( 1 + \frac{20}{9} \Psi \right) \delta_{ij}
            + 2 t^{\frac23} \Psi_{,ij}  \nonumber \\
          && + \,\frac{10}{9} t^{\frac23}
              \left( -6 \Psi_{,i}\Psi_{,j} - 4\Psi\Psi_{,ij}
                + \Psi^{,k}\Psi_{,k}\,\delta_{ij}
              \right)  \\
          && + \,\frac{1}{7} t^{\frac43}
            \left[ 19 \Psi^{,k}_{\ ,i}\Psi_{,kj}
                  -12 \Psi^{,k}_{\ ,k}\Psi_{,ij}
                  +3\left( \left(\Psi^{,k}_{\ ,k}\right)^2
                    - \Psi^{,k}_{\ ,\ell}\Psi^{,\ell}_{\ ,k} \right)
                   \delta_{ij}
            \right] \; . \nonumber
\end{eqnarray}
Therefore, we find that their ``fifth-order'' result coincides with our
second-order solution, except for the transverse-traceless part, $\chi_{ij}$.
If we take the long-wavelength limit, $\chi_{ij}$ can be neglected,
since spatial derivatives are assumed to be quantities of higher order
than time derivatives in this limit and as a result, the wave equation for
$\chi_{ij}$ does not appear. In this sense, our result includes
theirs.

\section{Concluding remarks}

In this paper, we have developed the second-order perturbative approach
to the nonlinear evolution of irrotational dust universes
in the framework of general relativity.
We have shown the complete calculation of the second-order solutions
in a $k = 0, \ \Lambda = 0$ background,
based on the tetrad formalism given by Kasai \cite{bi:kasai95}.
As mentioned in Sec. IV, our second-order solution
includes the results given in Tomita \cite{bi:tomita67},
Matarrese et al. \cite{bi:mata} and Parry et al. \cite{bi:pss}
although the essential calculation we need in our approach is
just the solution of a second-order ordinary differential equation
by iteration method. Therefore, our approach surpasses these others
in perfection and simplicity.

Another advantage of our method remains, which is not mentioned above.
Tomita's approach is valid only when the absolute value
of the density contrast $|\delta| \ll 1$ while ours does not rely
on this assumption, which is the inherent usefulness of
the so-called Zel'dovich approximation.
The gradient expansion technique implies taking the ``square root'' of the
metric
tensor in order to reproduce the Zel'dovich approximation,
while we do not need such a trick since we start from the tetrad formalism.

In our approach the extensions to
$k \ne 0, \ \Lambda \ne 0$ cases and radiation universes
$(p = \frac13 \rho)$ are straightforward.
These will be the subjects of future investigation.

\acknowledgments

H. R. wants to thank M. Soffel and H. Riffert for stimulating and helpful
discussions.
\newline
H. R. was supported by the ``Deutsche Forschungsgemeinschaft".
M. M. would like to thank T. Hamana for valuable comments.

\appendix
\section{Gauge condition}

The most general gauge transformation to first order is the result of
the infinitesimal coordinate transformation
\begin{equation}
\tilde x^{\mu}=x^{\mu}+\xi^{\mu} \;.
\end{equation}
The changes in the 4-velocity and in the metric tensor are computed from
\begin{equation}
\tilde u^{\mu}(\tilde x^{\lambda})=
\frac{\partial\tilde x^{\mu}}{\partial x^{\nu}}u^{\nu}(x^{\lambda})\;,
\qquad
g_{\mu\nu}(x^{\lambda})
=\frac{\partial \tilde x^{\alpha}}{\partial x^{\mu}}
 \frac{\partial \tilde x^{\beta }}{\partial x^{\nu}}
\tilde g_{\alpha\beta}(\tilde x^{\lambda})\;,
\end{equation}
which gives to first order
\begin{eqnarray}
\delta_G u^{\mu} &\equiv&
  \tilde u^{\mu}(x^{\lambda})-u^{\mu}(x^{\lambda})
=\xi^{\mu}_{\;,\nu}u^{\nu}-u^{\mu}_{\;,\nu}\xi^{\nu} \;, \\
\delta_G g_{\mu\nu} &\equiv&
  \tilde g_{\mu\nu}(x^{\lambda})-g_{\mu\nu}(x^{\lambda})
=-g_{\mu\nu,\alpha}\xi^{\alpha}
-g_{\mu\alpha}\xi^{\alpha}_{\;,\nu}-g_{\nu\alpha}\xi^{\alpha}_{\;,\mu}\;.
\nonumber
\end{eqnarray}

If the perturbations are linear, we can treat scalar perturbations seperate
and we write
\begin{equation}
\xi^{\mu} = (T, \delta^{ij}L_{,j})
\end{equation}
for the $k = 0$ background, where $T=T(x^{\mu})$ and
$L=L(x^{\mu})$ are arbitrary scalar functions.

The gauge condition we impose in this paper is the comoving synchronous
condition:
\begin{equation}
u^i=0 \;, \quad g_{00}=-1 \;, \quad g_{0i}=0 \;.
\end{equation}
These equations must hold for every gauge transformation,
so that $\delta_G u^i = \delta_G g_{00} = \delta_G g_{0i}=0$
lead to
\begin{equation}
\dot L_{,j}=0 \;,\quad \dot T=0 \;,\quad T_{,i}=0\;.
\end{equation}
Apart from a trivial constant translation, these are solved to give
\begin{equation}
T = 0 \;, \qquad L_{,j} = L_{,j}(\mbox{\bf x}) \;.
\end{equation}

The change due to the residual gauge freedom is
\begin{equation}
\delta_G g_{ij}=-2a^2 L_{,ij}(\mbox{\bf x}) \;,
\end{equation}
or if we use Eq.~(\ref{eq:gij1st}),
\begin{equation}
\delta_G C_{,ij}(\mbox{\bf x}) = -  L_{,ij}(\mbox{\bf x}) \;.
\end{equation}
Therefore, using the remaining gauge freedom $L_{,j}(\mbox{\bf x})$,
we can choose $C_{,ij}(\mbox{\bf x}) = 0$.

\section{Complete second-order solutions}

The complete solution for the triad reads
\begin{eqnarray}
  e^{(\ell)}_{\ i} = && \left(1 + \frac{10}{9} \Psi \right)
                                  \delta^{(\ell)}_{\ i}
     + \delta^{(\ell)}_{\ j} \left(
        t^{\frac{2}{3}} \;\Psi^{,j}_{\ ,i} + t^{-1} \;\Phi^{,j}_{\ ,i}
                             \right) \nonumber \\
   &&+ \ \delta^{(\ell)}_{\ j} \left(
        t^{\frac{2}{3}} \;\psi^j_{\ i} + t^{-1} \;\phi^j_{\ i}
      + t^{\frac{4}{3}} \;\varphi^j_{\ i}
      + t^{-\frac{1}{3}} \;\upsilon^j_{\ i} + t^{-2} \;\zeta^j_{\ i}
      + \chi^j_{\ i} + \vartheta^j_{\ i} + \theta^j_{\ i}
                             \right)  \; ,
\end{eqnarray}
where $\psi^i_{\ j}$ and $\phi^i_{\ j}$ are the spatially dependent parts
of the second-order homogeneous solutions of Eq.~(\ref{eq:key}),
$\varphi^i_{\ j}$, $\upsilon^i_{\ j}$ and $\zeta^i_{\ j}$
those of the second-order inhomogeneous solutions,
which come from $\Psi\times\Psi$, $\Psi\times\Phi$ and $\Phi\times\Phi$,
and $\chi^i_{\ j}$, $\vartheta^i_{\ j}$ and $\theta^i_{\ j}$
are the corresponding transverse-traceless parts.

\subsection{The coupling mode}

We obtain
\begin{eqnarray}
 \upsilon^i_{\ j}
  = && 2\Psi^{,k}_{\ ,k}\Phi^{,i}_{\ ,j} + 2\Phi^{,k}_{\ ,k}\Psi^{,i}_{\ ,j}
   -19 \Psi^{,i}_{\ ,k}\Phi^{,k}_{\ ,j} - 15\Psi^{,k}\Phi^{,i}_{\ ,jk} \\
    && + \left(\,
         6\Psi^{,k}_{\ ,\ell}\Phi^{,\ell}_{\ ,k}
       - \Psi^{,k}_{\ ,k}\Phi^{,\ell}_{\ ,\ell}
       + 5\Psi^{,k}\Phi^{,\ell}_{\ ,k\ell} \,
         \right) \delta^i_{\ j}
  + \frac{9}{2}\left(
      \phi^{k\ ,i}_{\ k\ ,j} - \phi^{i\ ,k}_{\ j\ ,k}
               \right)  \nonumber
\end{eqnarray}
with
\begin{equation}
  \phi^i_{\ j,i} - \phi^i_{\ i,j}
   = -\frac{20}{9} \Psi_{,k} \Phi^{,k}_{\ ,j} \; .
\end{equation}
(In this calculation we use
$\Psi^{,i}_{\ ,k}\Phi^{,k}_{\ ,j} = \Phi^{,i}_{\ ,k}\Psi^{,k}_{\ ,j}$ ,
which comes from
$V^i_{\ j} \equiv \frac12\gamma^{ik}\dot\gamma_{jk}
            = e^i_{(\ell)} \dot e^{(\ell)}_{\ j}$ .)

The equation for the transverse-traceless part is
\begin{equation}
 {\ddot\vartheta}^{i}_{\ j} + 3 \frac{\dot a}{a}{\dot\vartheta}^{i}_{\ j}
         - \frac{1}{a^2}\nabla^2{\vartheta^{i}_{\ j}}
  = t^{-\frac{5}{3}}{\cal P}^{i}_{\ j}\;,
\end{equation}
where
\begin{eqnarray}
  {\cal P}^i_{\ j}
    &=&  \left( 6\Psi^{,k}_{\ ,\ell}\Phi^{,\ell}_{\ ,k}
               - \Psi^{,k}_{\ ,k}\Phi^{,\ell}_{\ ,\ell}
               + 5\Psi^{,k}\Phi^{,\ell}_{\ ,k\ell}
         \right)^{,i}_{\ ,j} \nonumber \\
   && + \biggl[ \,2\Psi^{,k}_{\ ,k}\Phi^{,i}_{\ ,j}
                + 2\Phi^{,k}_{\ ,k}\Psi^{,i}_{\ ,j}
                - 24\Psi^{,i}_{\ ,k}\Phi^{,k}_{\ ,j}
                - 20\Psi^{,k}\Phi^{,i}_{\ ,jk} \\
    && + \left(
         6\Psi^{,k}_{\ ,\ell}\Phi^{,\ell}_{\ ,k}
       - \Psi^{,k}_{\ ,k}\Phi^{,\ell}_{\ ,\ell}
       + 5\Psi^{,k}\Phi^{,\ell}_{\ ,k\ell}
         \right) \delta^i_{\ j}
  + \frac{9}{2}\left(
      \phi^{k\ ,i}_{\ k\ ,j} - \phi^{i\ ,k}_{\ j\ ,k} \right)
       \,\biggr]^{,m}_{\ ,m} \; . \nonumber
\end{eqnarray}
Using the conformal time $\eta$, this is rewritten as
\begin{equation}
   \frac{\partial^2}{\partial\eta^2}\vartheta^i_{\ j}
 + \frac{4}{\eta}\frac{\partial}{\partial\eta}\vartheta^i_{\ j}
 - \nabla^2\vartheta^i_{\ j}
 = \frac{3}{\eta}{\cal P}^i_{\ j} \; .
\end{equation}
The solution is
\begin{equation}
 \vartheta^i_{\ j}(\mbox{\bf x},\eta)
   = \frac{3}{4\pi\eta^3}
      \int_0^{\eta-\eta_{in}} dr'\ r'\left(
        (\eta + r')(\eta - r') - r' \eta_{in}
                                     \right)
       \int d\Omega'\ {\cal P}^i_{\ j}(\mbox{\bf x} + \mbox{\bf x'}) \; .
\end{equation}

\subsection{The decaying mode}

We obtain
\begin{equation}
\zeta^i_{\ j} = \frac{1}{4} \left( \lambda^k_{\ k} \,\delta^i_{\ j}
                   -4 \lambda^i_{\ j} \right) \;,
\end{equation}
where
\begin{equation}
\lambda^i_{\ j} \equiv \frac{1}{2} \left(
  \Phi^{,k}_{\ ,k} \Phi^{,i}_{\ ,j} -
  \Phi^{,i}_{\ ,k} \Phi^{,k}_{\ ,j}\right) \;.
\end{equation}

The equation for the transverse-traceless part is
\begin{equation}
 \ddot\theta^i_{\ j} + 3 \frac{\dot a}{a}\dot\theta^i_{\ j}
      - \frac{1}{a^2}\nabla^2\theta^i_{\ j}
    = t^{-\frac{10}{3}}{\cal Q}^i_{\ j} \; ,
\end{equation}
where
\begin{equation}
{\cal Q}^i_{\ j} =
  \frac{1}{4} \lambda^{k\ ,i}_{\ k\ ,j}
 +\frac{1}{4} \left(\lambda^{k}_{\ k}\, \delta^i_{\ j} - 4 \lambda^i_{\ j}
              \right)\!{}^{,\ell}_{\ ,\ell} \; .
\end{equation}
Again this is rewritten as
\begin{equation}
   \frac{\partial^2}{\partial\eta^2}\theta^i_{\ j}
 + \frac{4}{\eta}\frac{\partial}{\partial\eta}\theta^i_{\ j}
 - \nabla^2\theta^i_{\ j}
 = \frac{729}{\eta^6}{\cal Q}^i_{\ j}
\end{equation}
and we obtain the solution
\begin{equation}
 \theta^i_{\ j}(\mbox{\bf x},\eta)
   = \frac{729}{16\pi\eta^3}
       \int_0^{\eta-\eta_{in}} dr'\ r'\left(
                         (4\eta-r')(\eta-r')^{-4}+r'\eta_{in}^{-4}\right)
        \int d\Omega'\ {\cal Q}^i_{\ j}(\mbox{\bf x}+\mbox{\bf x'}) \; .
\end{equation}

\subsection{The complete expression of the metric tensor}

The complete metric reads
\begin{eqnarray}
\gamma_{ij} = & &
  \left(1 + \frac{20}{9}\Psi \right) \delta_{ij}
      + 2t^{\frac{2}{3}} \Psi_{,ij} + 2t^{-1} \Phi_{,ij}\\
 && + \ \frac{10}{9}t^{\frac23}\left(
         - 6\Psi_{,i}\Psi_{,j} - 4\Psi\Psi_{,ij}
         + \Psi^{,k}\Psi_{,k}\,\delta_{ij}
                               \right)  \nonumber\\
 && + \ \frac17 t^{\frac43}\left[ \,19\Psi^{,k}_{\ ,i}\Psi_{,kj}
                  - 12\Psi^{,k}_{\ ,k}\Psi_{,ij}
                  + 3\left( \left(\Psi^{,k}_{\ ,k}\right)^2
                    - \Psi^{,k}_{\ ,\ell}\Psi^{,\ell}_{\ ,k}
                     \right) \delta_{ij}
                         \,\right] + 2 \chi_{ij}  \nonumber\\
 && + \ 2t^{-1}\left(
         \phi_{ij} + \frac{10}{9}\Psi\Phi_{,ij}
             \right) \nonumber\\
 && + \ t^{-2} \left[ \,2\Phi^{,k}_{\ ,i}\Phi_{,kj}
                     - \Phi^{,k}_{\ ,k}\Phi_{,ij}
        + \frac{1}{4} \left((\Phi^{,k}_{\ ,k})^2
                     - \Phi^{,k}_{\ ,\ell}\Phi^{,\ell}_{\ ,k}
                      \right)\delta_{ij}
             \,\right]    + 2 \theta_{ij} \nonumber\\
 && + \ t^{-\frac13} \biggl[ \,
          4\Psi^{,k}_{\ ,k}\Phi_{,ij} + 4\Phi^{,k}_{\ ,k}\Psi_{,ij}
       - 18\Psi^{,k}_{\ ,i}\Phi_{,kj} - 18\Phi^{,k}_{\ ,i}\Psi_{,kj}
       - 30\Psi_{,k}\Phi^{,k}_{\ ,ij}     \nonumber\\
 && \hspace{1cm}
       + \left(
           12\Psi^{,k}_{\ ,\ell}\Phi^{,\ell}_{\ ,k}
         - 2\Psi^{,k}_{\ ,k}\Phi^{,\ell}_{\ ,\ell}
         + 10\Psi^{,k}\Phi^{,\ell}_{\ ,k\ell}
         \right) \delta_{ij}
       + 9 \left(
              \phi^k_{\ k,ij} - \phi^{\ \ ,k}_{ij\ ,k}
           \right)
                   \, \biggr] + 2 \vartheta_{ij} \; .  \nonumber
\end{eqnarray}


\end{document}